# Phonon Dynamics and Inelastic Neutron Scattering of Sodium Niobate

S. K. Mishra[1], M. K. Gupta[1], R. Mittal[1], M. Zbiri[2], S. Rols[2], H. Schober[2] and S. L. Chaplot[1]

*Solid State Physics Division, Bhabha Atomic Research Centre, Trombay, Mumbai-400085, India*

*Institut Laue-Langevin, BP 156, 38042 Grenoble Cedex 9, France*

Sodium niobate ($NaNbO_3$) exhibits most complex sequence of structural phase transitions in perovskite family and therefore provides as excellent model system for understanding the mechanism of structural phase transitions. We report temperature dependence of inelastic neutron scattering measurements of phonon densities of states in sodium niobate. The measurements are carried out in various crystallographic phases of this material at various temperatures from 300 K to 1048 K. The phonon spectra exhibit peaks centered around 19, 37, 51, 70 and 105 meV. Interestingly, the peak around 70 meV shifts significantly towards lower energy with increasing temperature, while the other peaks do not exhibit an appreciable change. The phonon spectra at 783 K show prominent change and become more diffusive as compared to those at 303 K. In order to better analyze these features, we have performed first principles lattice dynamics calculations based on the density functional theory. The computed phonon density of states is found to be in good agreement with the experimental data. Based on our calculation we are able to assign the characteristic Raman modes in the antiferroelectric phase to the $A_{1g}$ symmetry, which are due to the folding of the T ($\omega$=95 cm$^{-1}$) and $\Delta$ ($\omega$=129 cm$^{-1}$) points of the cubic Brillouin zone.

**PACS**: 78.70.Nx, 63.20.-e, 63.20.dk

**Keywords:** Neutron inelastic scattering, Phonons in crystal lattice, First-principles DFT theory

# I. Introduction

Materials exhibiting ferroelectric/ piezoelectric properties are subject of keen interest due to their potentially practical applications ranging from high density memories to advanced robotic technology (as sensor and actuator) [1-4]. Niobate based materials are environment friendly and appropriate for wide piezoelectric applications due to their piezo-response that is comparable to Pb(Zr Ti)$O_3$. One of the end members, NaNb$O_3$ is a well-documented antiferroelectric that also finds applications in high density optical storage, enhancement of non-linear optical properties, as hologram recording materials, etc [3-5]. Relaxor type behavior in NaNb$O_3$ based solid solutions has also been reported [5].

Beyond the technological application, NaNb$O_3$ has been a rich model system for understanding of the mechanism of structural phase transitions. This system exhibits one of the most complex sequences of structural phase transitions in the perovskite family [6-7]. Above 913 K, it has a paraelectric cubic phase (**Pm$\bar{3}$m**). On lowering the temperature it undergoes transition to a series of antiferrodistortive phases in this order: tetragonal (*T*2) **P4/mbm**, orthorhombic (*T*1) **Cmcm**, orthorhombic (*S*) **Pbnm**, orthorhombic (*R*) **Pbnm**, orthorhombic (*P*) **Pbcm** phases, and a rhombohedral **R3c** phase. In a previous work, we have carried out detailed neutron diffraction study in the temperature range 17 to 350 K. We have evidenced unambiguously the coexisting of a ferroelectric (FE) **R3c** phase and an antiferroelectric (AFE) phase (**Pbcm**) over a wide range of temperatures. The coexistence of these phases and the reported anomalous dielectric response are consistent with competing ferroelectric and antiferroelectric interactions. Recent high pressure neutron diffraction measurements carried out up to 11 GPa at ambient temperature indicate transition from the **Pbcm** to the **Pbnm** phase [8]. These transitions are characterized by the appearance and the disappearance of superlattice reflections in the powder diffraction patterns. The superlattice reflections originate from the condensation of zone-centre and zone-boundary phonon modes. Thus, the stability of various crystallographic phases could be understood in terms of phonon instabilities.

The materials with perovskite structure have been subject of intense numerical investigations by means of first-principles calculations [9-13]. The focus is on the ground-state structure and properties, lattice dynamics, and dielectric and piezoelectric response functions. Virtually, all the perovskites exhibit high-symmetry (cubic) structure at high temperatures. Phonons have been known to play a key role in the understanding of structural phase transitions in ferroelectrics. Most of the ferroelectric transitions are governed by the softening of phonon modes in the high symmetry phase. First-principles calculation is known to be a valuable tool to identify phonon frequencies and other related characteristics. The cubic perovskite structure has unstable modes and it is therefore subject to energy-lowering distortions like zone-center distortions (resulting in ferroelectricity) and zone-boundary distortions involving rotations and/or tilting of the oxygen octahedra.

Phonons can be probed via various experimental methods, including infrared absorption, and inelastic neutron & light scattering [14-17]. To understand the phase transition behavior of NaNbO$_3$ some studies of the temperature dependence of the Raman and infrared spectra [15-19] were performed. However, these measurements are limited to the Brillouin zone centre and do not give a comprehensive picture of the dynamics. In this context, the combination of inelastic neutron scattering and first-principles calculations forms an adequate framework to obtain accurately phonon frequencies. Both the experimental and computational techniques are helpful to understand the role of structural distortions and their correlation to phonon instabilities, leading to phase transitions in this material. The aim of our study is to provide a better understanding of the phonon spectra of NaNbO$_3$ and to correlate the specific phonon modes relevant to the observed structural distortions. Inelastic neutron scattering provides an opportunity to validate our calculations. It provides important insights into the correlations between vibrational spectra and phase transition.

In the present study, we report results of inelastic neutron scattering measurements of phonon spectra in different crystallographic phases of sodium niobate. The interpretation and analysis of the observed phonon spectra and the identification of the specific phonons responsible for the stabilization of the ferroelectric and the antiferroelectric phases have been performed using ab-initio phonon calculations. This paper is organized as follows: the details of the experimental technique and first-principles calculations are summarized in section II and section III, respectively. Section IV is dedicated to the presentation and discussion of the results. Conclusions are drawn in section V.

## II. Experimental Details

The temperature dependent inelastic neutron scattering experiment on NaNbO$_3$ was carried out using the IN4C spectrometer at the Institut Laue Langevin (ILL), Grenbole, France. The spectrometer is based on the time-of-flight technique and is equipped with a large detector bank covering a wide range of about 10$^o$ to 110$^o$ of scattering angle. Polycrystalline sample of about 20 grams of NaNbO$_3$ was placed in a thin cylindrical niobium container for neutron measurements. The measurements were done in the neutron-energy-gain mode using the incident neutron energy of 14.2 meV (2.4 Å). Several inelastic runs were recorded on increase of temperature from 300 K to 1048 K. In the incoherent one-phonon approximation, the phonon density of states [20] is related to the measured scattering function $S(Q,E)$, as observed in the neutron experiments by:

$$g^{(n)}(E) = A < \frac{e^{2W_k(Q)}}{Q^2} \frac{E}{n(E,T) + \frac{1}{2} \pm \frac{1}{2}} S(Q,E) > \qquad (1)$$

$$g^n(E) = B \sum_k \{\frac{4\pi b_k^2}{m_k}\} g_k(E) \qquad (2)$$

where the + or − signs correspond to energy loss or gain of the neutrons, respectively, and $n(E,T) = [\exp(E/k_B T) - 1]^{-1}$. *A* and *B* are normalization constants and $b_k$, $M_k$, and $g_k(E)$ are, respectively, the neutron scattering length, mass, and partial density of states of the *k* th atom in the unit cell. The quantity within angular brackets represents the average over all *Q* values. $2W_k(Q)$ is the Debye-Waller factor.

### III. Computational Details

Relaxed geometries and total energies were obtained using the projector-augmented wave formalism [21] of the Kohn-Sham formulation of the density-functional theory [22] at the generalized gradient approximation (GGA) level, implemented in the Vienna *ab initio* simulation package (VASP) [23]. The GGA was formulated by the Perdew-Burke-Ernzerhof (PBE) density functional [24]. All results are well converged with respect to *k* mesh and energy cutoff for the plane-wave expansion. The total energy calculations have been done using an energy cutoff of 1100 eV. A grid of 8×8×8 K-points was used according to the Monkhorst-Pack (MP) scheme [25]. The break conditions for the self-consistent field and for the ionic relaxation loops were set to $10^{-8}$ eV Å$^{-1}$ and $10^{-4}$ eV Å$^{-1}$, respectively. Hellmann-Feynman forces following geometry optimization were less than $10^{-4}$ eV Å$^{-1}$. Full geometry optimization, including cell parameters, was carried out on NaNbO$_3$, under the phases AFE and FE. Neutron diffraction measurements showed that the two phases coexist over a wide temperature range (10- 250 K) [6]. Table I compares the calculated and experimental structural parameters for both the phases. The calculated lattice parameters are found to be slightly overestimated as compared to the experimental ones, as expected from GGA calculations.

At room temperature NaNbO$_3$ crystallizes in the orthorhombic phase (***Pbcm***) with 8 f.u. per unit cell (40 atoms). This leads to 120 phonon branches (3 acoustic modes + 117 optical modes). From the group theoretical analysis, the irreducible representations of the zone-centre optical phonons are $\Gamma_{optical}$ = **15A$_g$ + 17B$_{1g}$ + 15B$_{2g}$ + 13B$_{3g}$ + 13A$_u$ + 14B$_{1u}$ + 16B$_{2u}$ + 14B$_{3u}$,** where the $A_g$, $B_{1g}$, $B_{2g}$ and $B_{3g}$ modes are Raman active, whereas the modes $A_u$ are both Raman and infrared inactive. The modes $B_{1u}$, $B_{2u}$ and $B_{3u}$ are infrared active. Hence, 60 Raman active modes are expected in the orthorhombic phase (***Pbcm***).

At low-temperature NaNbO$_3$ is rhombohedral *(R3c)* with 2 f.u. per unit cell (10 atoms), resulting in 30 phonon branches (3 acoustic modes + 27 optical modes). The irreducible representations of the zone-centre optical modes are: $\mathbf{\Gamma_{optical} = 5A_1 + 4A_2 + 9E}$. The $A_1$ and the doubly degenerate *E* modes are both Raman and infrared active, whereas the $A_2$ mode is both Raman and infrared inactive. Therefore, 23 Raman active phonons are expected in the rhombohedral phase *(R3c)*.

In order to determine all force constants, the supercell approach was used for the lattice dynamics calculations. Total energies and Hellmann-Feynman forces were calculated for 42, 18 and 8 structures corresponding to the ***Pbcm, R3c*** and ***Pm-3m*** phases, respectively, and resulting from individual displacements of the symmetry inequivalent atoms in the supercell, along the inequivalent Cartesian directions ((±x, ±y, ±z). Phonon frequencies, Raman modes and dispersion relations were extracted from subsequent calculations using the direct method [26] as implemented in the PHONON software [27]. The computed zone centre phonon frequencies for both the AFE and FE phases are shown in Figure 1. As expected from the group theoretical analysis, the ferroelectric phase exhibits less Raman lines than it is found in the antiferroelectric phase.

## IV. Results and Discussion

Figure 2 depicts the evolution of inelastic neutron scattering (INS) spectra for NaNbO$_3$ at T= 303, 783, 838, 898, and 1048 K. The spectra correspond to different crystallographic phases. Five features (broad peaks centered around 19, 37, 51, 70 and 105 meV) can be easily identified. At 303 K, additional well resolve peaks below 37 meV are also observed. Their intensity decreases significantly with increasing temperature. Remarkably, the peak around 70 meV, shifts significantly towards lower energies with increasing temperature, while the others do not change in a noticeable way. At 783 K, a prominent change is observed and the spectra become more diffusive than that at 303 K. The variation of the INS spectra is associated with the occurrence of phase transitions.

Figure 3 compares the experimental and calculated neutron-weighted phonon density of states of NaNbO$_3$, in the antiferroelectric phase. The calculations are found to be in a fair agreement with the measurements. All the observed features are computationally well reproduced. Our lattice dynamics calculations show that both the ferroelectric (***R3c***) and the antiferroelectric (***Pbcm***) phases are dynamically stable. Further, from the calculated free-energies it is also found that the FE phase is energetically more favorable than AFE phase, with a tiny energy difference. This would explain the nature of both the phases in terms of their coexistence.

The ab-inito derived atomistic partial densities of states are shown in Figure 4. The oxygen atoms contribute dynamically in the whole spectral range, upto 120 meV, while Nb atoms mainly contribute upto 75 meV. The vibrations due to Na atom extend upto 40 meV. Computed partial density of state of Na reveals the presence of three peaks in antiferroelectric phase and two peaks for ferroelectric phase. It can also be seen that spread in the partial density of state associated with Na is more in AFE phase as compared to FE phase. It can be interpreted in terms of Na-O bond lengths. For the antiferroelectric phase, there are two types of sodium and four types of oxygen atoms, which result in significantly variation in bond length (change the force constant), and in turn result in more spread in the partial density of state. While in the ferroelectric phase, we have only one type of Na and O atoms. In the ferroelectric phase Na atoms are shifted from centre of the oxygen cage results and results in two different Na-O bond lengths. Similar interpretation also holds for partial density of states of niobium and oxygen atoms. It should be noticed that shorter bond lengths in FE phase (w.r.t AFE phase) results is the extending the energy range of the total density of state. It is interesting to note that band gap in the phonon density of states for FE phase is larger as compare to that in AFE phase.

To understand the vibrational properties of sodium niobate using inelastic scattering techniques various attempts have been taken [15-19]. However they were limited to the zone centre only. To the best of our knowledge, there are no experimental or theoretical reports extending the work to the zone boundaries (phonon dispersion relations) for antiferroelectric phase. This might be due to the experimental complexity associated with measuring the 120 modes. The synthesis of a suitable single domain crystal could add more difficulties as well. Presently, we report our ab- initio calculated phonon-dispersion relations (Figure 5). These are plotted along the high symmetry directions of the Brillouin zone. Figure 5 contains also available Raman data from the literature. Our calculations are in a good agreement with the available zone-centre experimental data.

The structural phase transitions in Perovskite-type materials ($ABO_3$) originate from the competing interactions between different phonon instabilities occurring in the cubic phase. These transitions belong generally to two classes: ferrodistortive (FD) and antiferrodistortive (AFD) [1, 28]. The FD and AFD phase transitions are driven by zone centre (q=0) and zone- boundary phonons (q≠ 0), respectively. Known examples of these transitions are the cubic to tetragonal phase transition in $BaTiO_3$ and $PbTiO_3$, for the ferrodistortive case, and in $SrTiO_3$ and $CaTiO_3$ for the antiferrodistortive case [1, 28]. The evolution of these phase transitions depends on the condensation sequence of the soft modes $M_3$ and $R_{25}$. The zone boundary $R_{25}$ mode is three fold degenerate and the $M_3$ mode is nondegenerate. The triply degenerate $R_{25}$ soft mode is made up of three components corresponding to the rotational degrees of freedom of the $BO_6$ octahedra around the three separate [001] axes. If one of the components condenses at the transition point, the resulting structure would be tetragonal *I4/mcm*, and the coupled condensation of the three components would lead to a rhombohedral *R3c* structure. However, when successive phase transitions associated with both the $M_3$ and $R_{25}$ soft modes, the sequence of the phase evolution depends in a complex way on the condensation sequence of the soft mode [1, 28].

Recently, Izumi and coworkers performed a detailed inelastic neutron scattering study in the cubic phase of $NaNbO_3$ [14]. Their measurements show gradual softening of the transverse acoustic (TA) phonon modes at the zone boundary points M (½ ½ 0) and R (½ ½ ½). This indicates instabilities of the in-phase and out-of-phase rotations of the oxygen octahedra around the [001] direction. The softening of these modes suggests low-lying flat transverse acoustic dispersion relations along the zone-boundary line M-R (T-line). As the temperature is decreased, these modes soften and become stable below the phase transition temperature. In order to detect these phonon instabilities using first principle technique, we have calculated the phonon dispersions from the zone centre ($\Gamma$) to the zone boundary points R and M (Figure 6). These are compared to experimental data from the literature [14]. The overall agreement is found to be satisfactory. Small deviations are expected as calculations were obtained at 0 K whereas inelastic neutron scattering data were acquired at in cubic phase (970 K).

From Figure 6 one can see that, in contrast to $SrTiO_3$ [29], the polar instability strength at the $\Gamma$ point is stronger than the antiferrodistortive instabilities at the R and M points, and it extends over a wider region of the Brillouin zone. Further, the strength of the M and the R point instabilities are quite similar. Interestingly, the branches along the $\Gamma$-R and $\Gamma$-M directions, show dramatic changes when reaching the R and M points. When moving away from M to R, two unstable modes are detected. One of them is rather flat and the other one shows rapid stiffening and becomes stable. Moreover, one of the stable modes become unstable at **T** (½ ½ ¼) point. Our results are consistent with other theoretical works in the literature.

Above 950 K, $NaNbO_3$ occurs in the cubic phase. On decreasing the temperature, it transforms to a tetragonal phase (***P4/mbm***). The first structural transformation is from cubic to tetragonal structure, where the unit cell is doubled in the plane perpendicular to the rotation axes of the $M_3$ mode. By further lowering the temperature, condensation of the $R_{25}$ phonon leads to the orthorhombic ***Cmcm*** ($T_2$) phase. Unstable phonon-branches along the M-R line contribute to the occurring successive phase transitions. These phonons play an important role in stabilizing the different phases (P, S and R) in $NaNbO_3$. In a previous study [7], we have proposed that the additional superlattice reflections are due to the condensation of the zone boundary phonons at T (**q**= ½, ½, g). The orthorhombic structures of the S, R and P phases result from the condensation of the phonon mode (q= ½, ½, g); with g= 1/12, 1/6 and 1/4. These orthorhombic phases originate from the modulation of the high symmetry cubic phase, associated with the phonon modes at **q**= (½, ½, g). Further, the freezing of all the $R_{25}$ modes and a zone-centre phonon stabilizes the low-temperature ferroelectric rhombohedral phase.

Deeper insights into phonon dynamics can be gained by performing an analysis of the eigenvectors corresponding to specific phonon modes, relevant to the present study. These are derived from our ab-initio calculations and are plotted in Figure 7. The eigenvector of the unstable Γ-point zone-centre phonon mode at $\omega = 23i$ meV indicates clearly that niobium and oxygen atoms are moving in opposite directions. This leads to the formation of a dipole, and induces ferroelectricity. The eigenvectors corresponding to the unstable modes at M and R points ($\omega=14i$ meV) exhibit an in-phase and out phase rotation of the oxygen octahedra, leading to a doubling of the unit cell. The analysis of the eigenvector of the X-point zone boundary mode at $\omega=15i$ meV suggests that, similarly to the zone centre mode, Nb and O atoms move in opposite directions within a layer of the basal plan, and this motion is antiphased in an adjacent layer. This results in a zero total dipole moment in unit cell. The displacement patterns are therefore related to antiferroelectricity. The mode at the T-point having the phonon frequency $\omega=14i$ meV possesses an eigenvector displacement indicating a multiplication of the unit cell.

Before we close, we would like to identify the phonons which are responsible for the stabilization of antiferroelectric phase. The AFE phase is found to accompany new super lattice reflections in powder neutron diffraction data [6-7]. This is confirmed by the appearance of new Raman lines in Raman spectroscopy. These lines become active due to the folding of the corresponding specific zone-centre points, below the AFE phase transition temperature. To investigate these special points, we have established a symmetry-based correlation between the orthorhombic zone centre points and the high symmetry points in the cubic phase (Table 2). Below the AFE phase transition, strong modifications of the Raman scattering patterns are observed [15-19] accompanying the appearance of new Raman modes around 93 $cm^{-1}$ and 123 $cm^{-1}$. Further, a sudden enhancement of the intensity of the bands within the two frequency ranges 150-300 $cm^{-1}$ and 500-650 $cm^{-1}$ are also noted [15-19]. We have assigned the two lines at 93 $cm^{-1}$ and 123 $cm^{-1}$ as belonging to the $A_{1g}$ irreducible representation, and we have identified them as originating from the folding of the T (93 cm$^{-1}$) and Δ (129 cm$^{-1}$) points of the Brillouin zone under the cubic phase. The eigenvectors corresponding to the two AFE modes of NaNbO$_3$, as extratcted from our ab initio lattice dynamical calculations are shown in Figure 8. The mode at 93 cm$^{-1}$ involves significant motions of Na, Nb and O, which are located at the sites 4d (¼+u, ¾+v, ¼), 8e (¼+u, ¼+v, ⅛+w) and ( ½ +*u*,0+*v*, ⅛+*w)*, respectively. However, the Raman mode at 129 cm$^{-1}$ reflects a significant displacement of all the atoms.

## V. Summary

We have reported inelastic neutron scattering (INS) measurements of the phonon density of states of sodium niobate as a function of temperature. The INS spectra are correlated to the various crystallographic phases of NaNbO$_3$, and show significant changes with increasing temperature. The peak at 70 meV is found to shift to lower energies, while the other pronounced peaks are not noticeably affected. Upon heating, the spectra become more diffusive. In order to get deeper insights into the observed features, we have performed ab- initio lattice dynamics calculations. The computed phonon density of states of NaNbO$_3$ is found to be in good agreement with our INS measurements. The calculated partial densities of states reveal that the dynamical contribution of the oxygen atoms spreads over the whole energy range, while the Nb atoms contribute mainly up to 75 meV. The vibrations due to Na atoms extend up to 30 meV. Above 75 meV, the dynamics is mainly due to the Nb-O stretching modes. Using the calculation, we have identified the various soft phonon modes at specific points in the Brillouin zone that are associated with various phase transition as a function of temperature. Further, we have found that the characteristic antiferroelectric Raman modes, which appear below the antiferroelectric phase transition temperature, correspond to the A$_{1g}$ symmetry and are due to the folding of the T (ω=95 cm$^{-1}$) and Δ (ω=129 cm$^{-1}$) points of the Brillouin zone, under the cubic phase.

**Acknowledgments**

We would like to thank Mrs. P. Goel, Solid State Physics Division, Bhabha Atomic Research Centre, Mumbai, India for her cooperation during inelastic neutron measurements at ILL.

1. M. E. Lines and A. M. Glass "Principles and Application of Ferroelectrics and Related Materials" (oxford: Clarendon, 1977); Xu. Yuhuan, Ferroelectric Materials and Their Applications (Nort-Holland Elsevier Science, 1991); L. G. Tejuca and J. L. G. Fierro "Properties and Applications of Perovskite-Type Oxides" (New York: Dekker, 1993).
2. E. Bousquet, M. Dawber, N. Stucki, C. Lichtensteiger, P. Hermet, S. Gariglio, J. M. Triscone and P. Ghosez, Nature **452**, 723 (2008); T. Choi, Y. Horibe, H. T. Yi, Y. J. Choi,WeidaWu and S.-W. Cheong, Nature materials **9**, 253 (2010).
3. Y. Saito, H. Takao, T. Tani, T. Nonoyama, K. Takatori, T. Homma, T. Nagaya, M. Nakamura, Nature **423**, 84 (2004); E. Cross, Nature **432**, 24 (2004).
4. E. Valdez, C. B de Araujo, A. A. Lipovskii Appl. Phys. Lett., **89**, 31901 (2006); E Hollenstein, M. Davis, D. Damjanovic and Nava Setter Appl. Phys. Lett., **87**, 182905 (2006); Yu. I. Yuzyuk, P. Simon, E. Gagarina, L. Hennet, D. Thiaudiere, V. I. Torgashev, S. I. Raevskya, I. P. Raevskii, L. A. Reznitchenko and J. L. Sauvajol, J. Phys.: Condens. Matter **17**, 4977 (2005).
5. P. Raevski, S. I. Raevskaya, S. A. Prosandeev, V. A. Shuvaeva, A. M. Glazer, and M. S. Prosandeeva, J. Phys.: Condens. Matter **16**, L221 (2004).
6. S. K. Mishra, N. Choudhury, S. L. Chaplot, P. S. R. Krishna and R. Mittal, Phys. Re**v B 76,** 024110 (2007) and reference therein.
7. S. K. Mishra, R. Mittal, V. Y. Pomjakushin and S. L. Chaplot, Phys. Rev **B 83,** 134105 (2011).
8. S. K. Mishra, M. K. Gupta, R. Mittal, S. L. Chaplot and T Hansen Appl. Phys. Lett. **101** 242907 (2012).
9. R Machado, M Sepliarsky and M G Stachiotti, Phys. Rev **B** 84, 134107 (2011).
10. W. Zhong and D. Vanderbilt, Phys. Rev. Lett. **74,** 2587 (1995).
11. O. Diéguez, K. M. Rabe and D. Vanderbilt, Phys Rev. B **72,** 144101 (2005).
12. W. Zhong, R. D. King-Smith and D. Vanderbilt, Phys. Rev. Lett. **72**, 3618 (1994).
13. R. D. King-Smith and D. Vanderbilt, Phys. Rev. B **49**, 5828 (1994).
14. I. Tomeno, Y. Tsunoda, K. Oka, M. Matsuura and M. Nishi, Phys Rev. B **80**, 104101 (2009).
15. S. J. Lin, D. P. Chiang, Y. F. Chen, C. H. Peng, H. T. Liu, J. K. Mei, W. S. Tse, T.-R. Tsai, and H.-P. Chiang, J. Raman Spectrosc. **37**, 1442 (2006).
16. E. Bouizane, M. D. Fontana, and M. Ayadi, J. Phys.: Condens. Matter **15**, 1387 (2003).
17. R. J. C. Lima, P. T. C. Freire, J. M. Saski, A. P. Ayala, F. E. A. Melo, J. Mendes Filho, K. C. Serra, S. Lanfredi, M. H. Lente, and J. A. Eiras, J. Raman Spectrosc. **33**, 669 (2002).
18. Z. X. Shen, X. B. Wang, S. H. Tang, M. H. Kuok, and R. Malekfar, J. Raman Spectrosc. **31**, 439 (2000). Z. X. Shen, X. B. Wang, M. H. Kuok, and S. H. Tang, J. Raman Spectrosc. **29**, 379 (1998). X. B. Wang, Z. X. Shen, Z. P. Hu, L. Qin, S. H. Tang, and M. H. Kuok, J. Mol. Struct. **385**, 1 (1996).

19. Y. Shiratori, A. Magrez, M. Kato, K. Kasezawa, C. Pithan, R. Waser, J. Phys. Chem. C **112**, 9610 (2008); Y. Shiratori, A. Magrez, J. Dornseiffer, F. H. Haegel, C. Pithan, R. Waser, J. Phys. Chem. B. **43**, 20122 (2005); Y. Shiratori, A. Magrez, W. Fischer, C. Pithan, and R. Waser**, 111**, 18493 (2007).
20. D. L. Price and K. Skold, in Neutron Scattering , edited by K. Skold and D. L. Price (Academic Press, Orlando, 1986), Vol. A; J. M. Carpenter and D. L. Price, Phys. Rev. Lett. **54**, 441 (1985).
21. P. E. Blöchl, Phys. Rev. B **50**, 17953 (1994);
22. P. Hohenberg and W. Kohn, Phys. Rev. **B 136**, 864 (1964); W. Kohn and L. J. Sham, Phys. Rev. **140**, A1133 (1965).
23. G. Kresse and J. Furthmüller, Comput. Mater. Sci. **6**, 15 (1996); G. Kresse and D. Joubert, Phys. Rev. B **59**, 1758 (1999).
24. J. P. Perdew, K. Burke, and M. Ernzerhof, Phys. Rev. Lett. **77**, 3865 (1996); J. P. Perdew, K. Burke, and M. Ernzerhof, Phys. Rev. Lett. **78**, 1396 (1997).
25. H. J. Monkhorst and J. D. Pack, Phys. Rev. B **13**, 5188 (1976).
26. K. Parlinski, Z.-Q. Li, and Y. Kawazoe, Phys. Rev. Lett. **78**, 4063 (1997).
27. K. Parlinski, software PHONON, 2010.
28. A. D. Bruce and R. A. Cowely, Adv. Phys. **29**, 219 (1980).
29. N. Choudhury, E. J. Walter, A. I. Kolesnikov, and C.-K. Loong Phys. Rev. B **77**, 134111 (2008).

**Table-1:** Experimental [6] and ab-inito calculated structural parameters of $NaNbO_3$ in the orthorhombic, antiferroelectric phase (***Pbcm***) and in the rhombohedral, ferroelectric phase (***R3c***).

| Orthorhombic Antiferroelectric phase (***Pbcm***) | | | | | | |
|---|---|---|---|---|---|---|
| | Experimental positional coordinates | | | Calculated positional coordinates | | |
| Atoms | x | y | z | x | y | z |
| Na1 | 0.247 | 0.75 | 0.00 | 0.260 | 0.7500 | 0.000 |
| Na2 | 0.227 | 0.789 | 0.25 | 0.259 | 0.796 | 0.250 |
| Nb | 0.242 | 0.282 | 0.131 | 0.243 | 0.279 | 0.125 |
| O1 | 0.329 | 0.25 | 0.00 | 0.310 | 0.250 | 0.00 |
| O2 | 0.208 | 0.278 | 0.25 | 0.191 | 0.227 | 0.25 |
| O3 | 0.530 | 0.040 | 0.138 | 0.542 | 0.033 | 0.140 |
| O4 | 0.975 | 0.489 | 0.107 | 0.960 | 0.456 | 0.110 |
| | Lattice Parameters (Å) $A_{orth}$= 5.5012 (Å); $B_{orth}$ = 5.5649 (Å), $C_{orth}$= 15.3972 (Å) | | | Lattice Parameters (Å) $A_{orth}$= 5.568 (Å); $B_{orth}$= 5.645 (Å), $C_{orth}$= 15.603 (Å) | | |

| Ferroelectric Rhombohedral phase (***R3c***) | | | | | | |
|---|---|---|---|---|---|---|
| | Experimental positional coordinates | | | Calculated positional coordinates | | |
| Atoms | x | y | z | x | y | z |
| Na | 0.272 | 0.272 | 0.272 | 0.273 | 0.273 | 0.273 |
| Nb | 0.016 | 0.016 | 0.016 | 0.014 | 0.014 | 0.014 |
| O | 0.320 | 0.183 | 0.747 | 0.312 | 0.184 | 0.749 |
| | Lattice Parameters (Å) $a_{rhom}$= 5.552 (Å) | | | Lattice Parameters (Å) $a_{rhom}$=5.637 (Å) | | |

**Table 2.** Correlation diagram between specific symmetry-points in the cubic phase and the zone-centre irreducible representations of the antiferroelectric, orthorhombic phase.

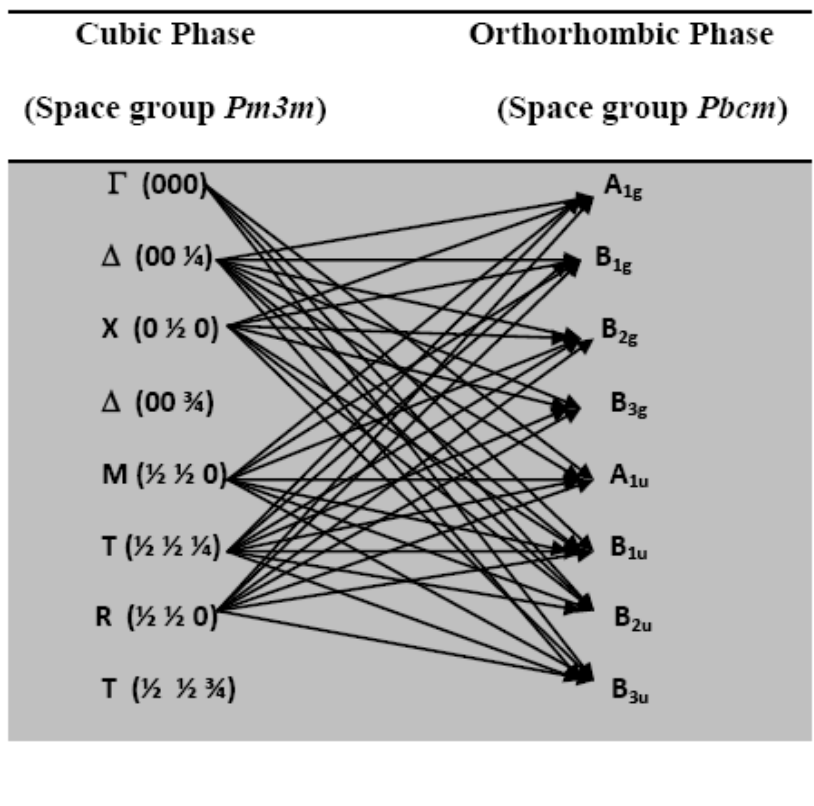

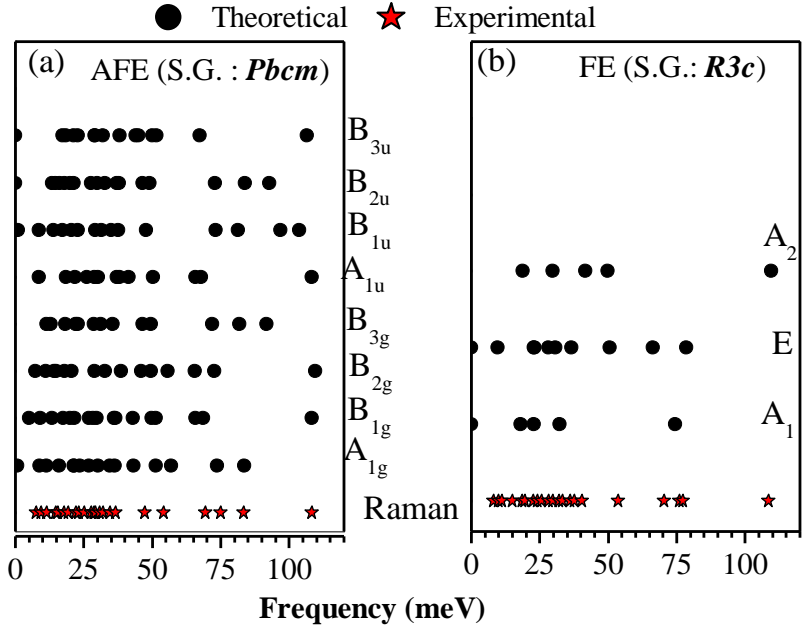

**Fig 1.** Comparison of the calculated (filled circles) long-wavelength phonon frequencies with the available experimental data (stars) [Ref.15-18] for both the antiferroelectric (AFE) and the ferroelectric (FE) phases.

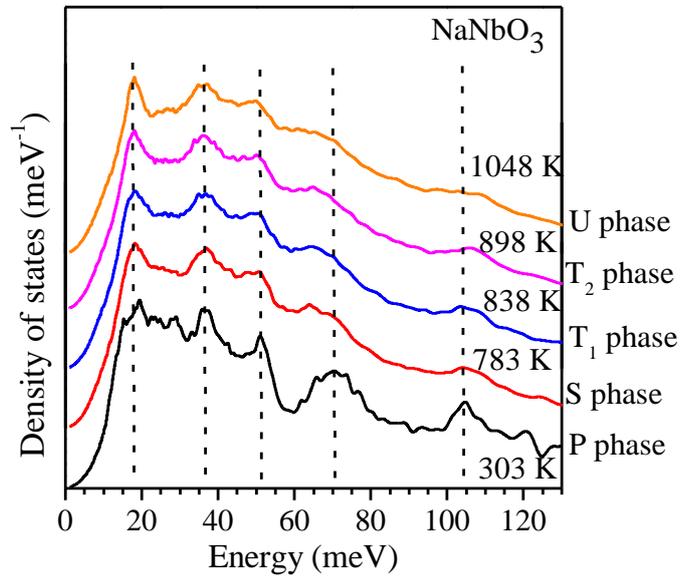

**Fig 2.** The temperature dependence of the phonon spectra of $NaNbO_3$ as observed by neutron inelastic scattering.

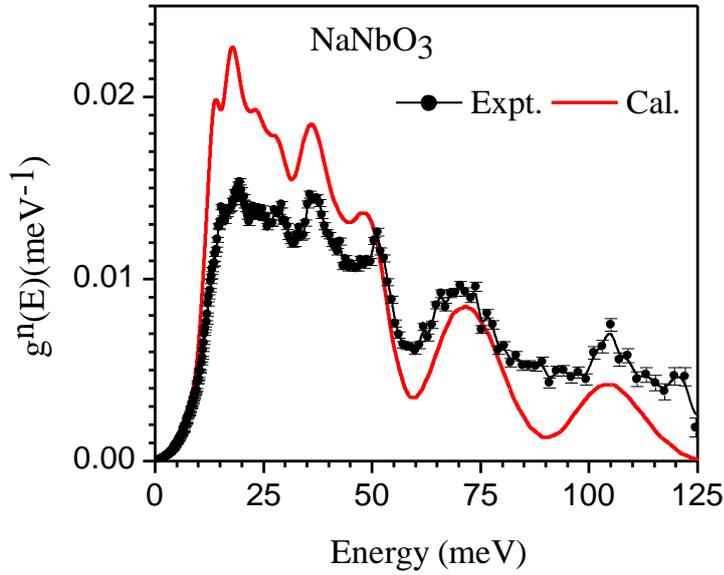

**Fig 3.** The experimental (dotted line at 303 K) and calculated (solid line at 0 K) phonon spectra for NaNbO$_3$ in the antiferroelectric phase (***Pbcm***). The calculated spectra have been convoluted with a Gaussian of FWHM of 15% of the energy transfer in order to describe the effect of energy resolution in the experiment.

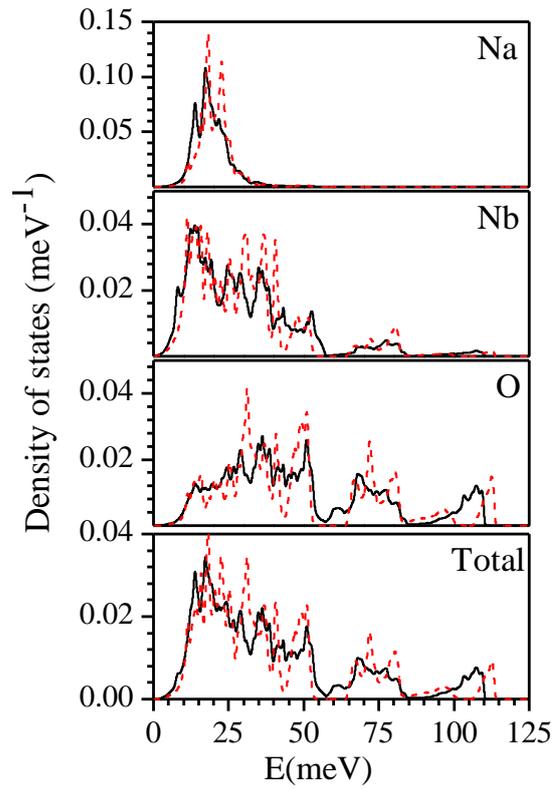

**Fig. 4:** The calculated partial density of states for various atoms and the total phonon density of states for NaNbO$_3$, in both the antiferroelectric orthorhombic (***Pbcm***) phase (solid line) and the ferroelectric rhombohedral (***R3c***) phase (dashed line).

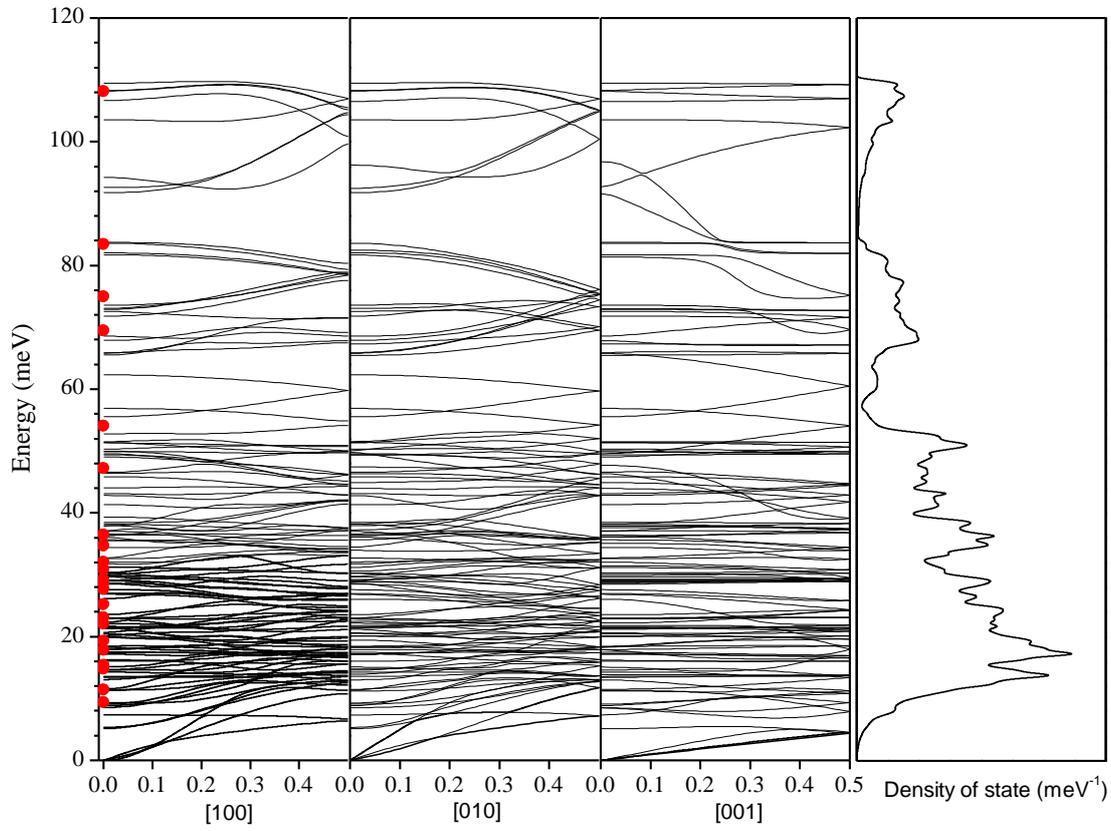

**Fig 5:** The calculated phonon dispersion relations for the orthorhombic phase (***Pbcm***) of NaNbO$_3$. For the sake of comparison, available experimental optical long wavelength data are also shown (solid circles) [Ref 15-18].

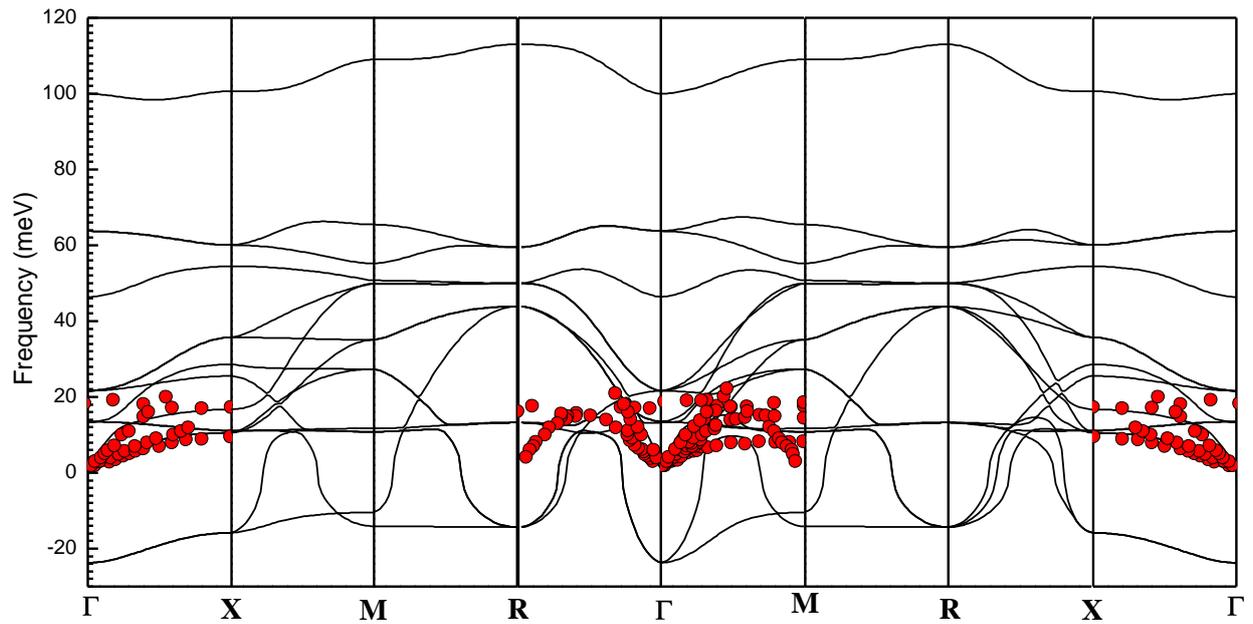

**Fig 6:** Computed phonon dispersion relations for cubic phase (*Pm-3m*) of NaNbO$_3$ compared to reported experimental inelastic neutron scattering (INS) single crystal data (red circle) (Ref. 14).

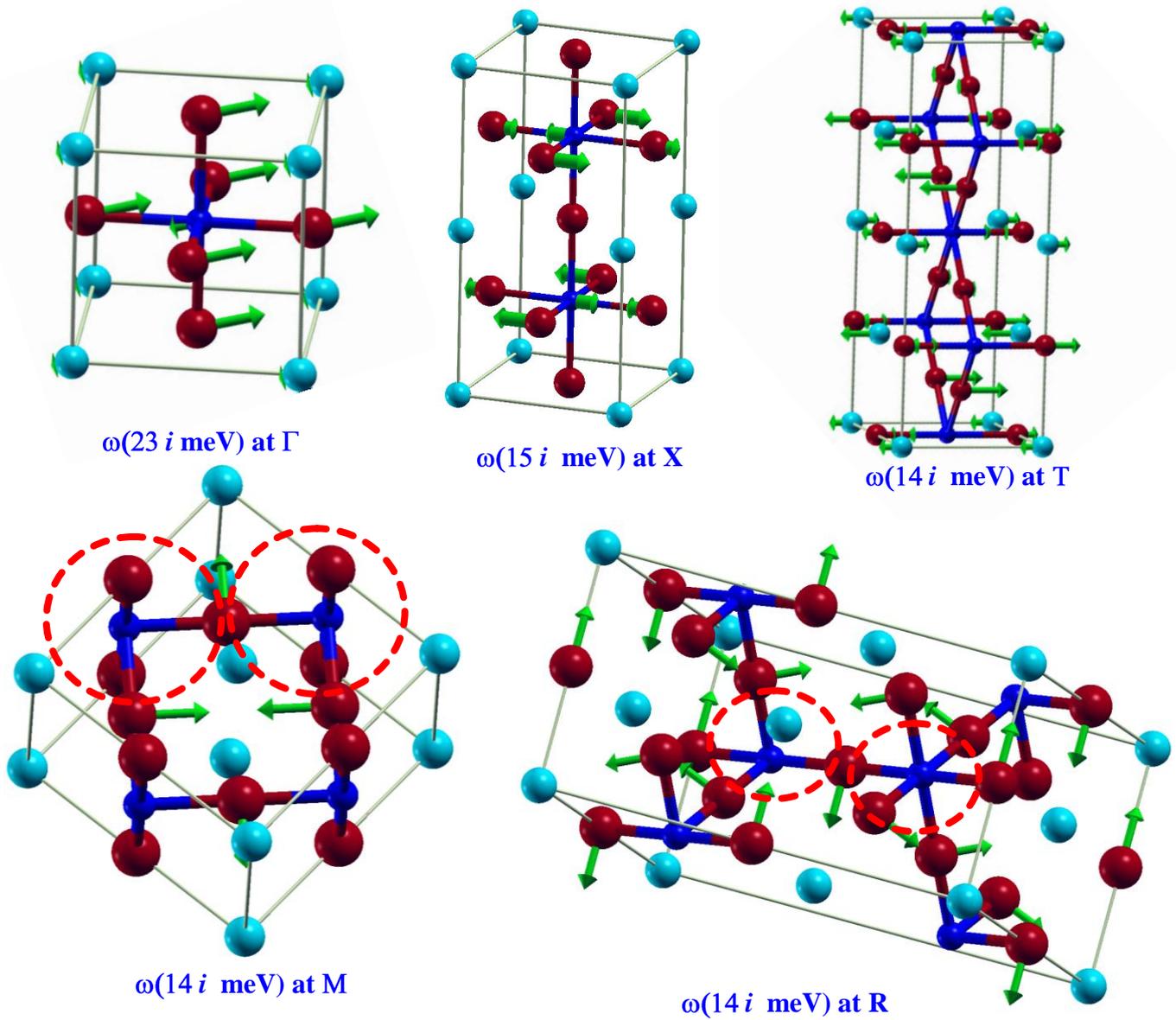

**Fig 7:** Ab initio derived eigenvectors of selected zone-centre an zone-boundary unstable phonon modes at the Γ, M, R, T and X points for the cubic phase of NaNbO$_3$. The lengths of arrows are related to the displacements of the atoms. Key: Na, cyan; Nb, blue; O, brown.

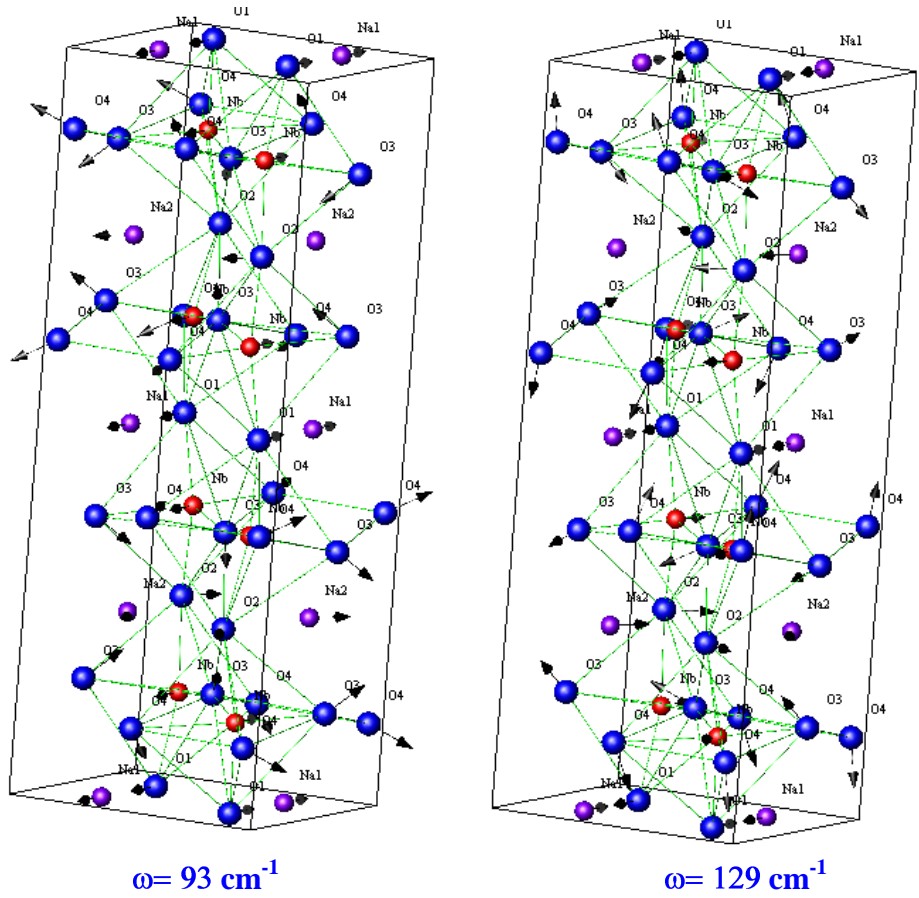

ω= 93 cm$^{-1}$             ω= 129 cm$^{-1}$

**Fig. 8.** (Color online) The eigenvectors of the two antiferroelectric modes, at (a) ω=93 cm$^{-1}$ and (b) 129 cm$^{-1}$ of NaNbO$_3$, induced by the folding of the T ( q= ½ ½ ¼ ) and Δ ( q= 0 0 ¼ ) points of the Brillouin zone under the cubic phase, respectively. The lengths of arrows are related to the displacements of the atoms. (Key: Na: violet spheres; Nb: blue spheres; O: brown spheres).